\documentclass[prd,twocolumn,showpacs,superscriptaddress]{revtex4-1}
\usepackage{graphicx}
\usepackage{units}
\usepackage{multirow}
\usepackage{bigstrut}
\usepackage{overpic}
\usepackage{array}
\usepackage{color}
\usepackage{amsmath}
\usepackage{bm}
\usepackage{times}
\RequirePackage{lineno}
\begin{document}

\modulolinenumbers[2]

\setlength{\oddsidemargin}{-0.5cm} \addtolength{\topmargin}{15mm}

\title{\boldmath Determination of strong-phase parameters in $D\rightarrow K^0_{S,L}\pi^+\pi^-$ }

\author{
  \small
M.~Ablikim$^{1}$, M.~N.~Achasov$^{10,d}$, P.~Adlarson$^{59}$, S. ~Ahmed$^{15}$, M.~Albrecht$^{4}$, M.~Alekseev$^{58A,58C}$, D.~Ambrose$^{51}$, A.~Amoroso$^{58A,58C}$, F.~F.~An$^{1}$, Q.~An$^{55,43}$
% ...
      \\
      \vspace{0.2cm}
      (BESIII Collaboration)\\
      \vspace{0.2cm} {\it
$^{1}$ Institute of High Energy Physics, Beijing 100049, People's Republic of China\\
% ...
$^{4}$ Bochum Ruhr-University, D-44780 Bochum, Germany\\
% ...
$^{10}$ G.I. Budker Institute of Nuclear Physics SB RAS (BINP), Novosibirsk 630090, Russia\\
% ...
$^{15}$ Helmholtz Institute Mainz, Johann-Joachim-Becher-Weg 45, D-55099 Mainz, Germany\\
% ...
$^{43}$ State Key Laboratory of Particle Detection and Electronics, Beijing 100049, Hefei 230026, People's Republic of China\\
% ...
$^{51}$ University of Minnesota, Minneapolis, Minnesota 55455, USA\\
% ...
$^{55}$ University of Science and Technology of China, Hefei 230026, People's Republic of China\\
% ...
$^{58}$ (A)University of Turin, I-10125, Turin, Italy; (B)University of Eastern Piedmont, I-15121, Alessandria, Italy; (C)INFN, I-10125, Turin, Italy\\
$^{59}$ Uppsala University, Box 516, SE-75120 Uppsala, Sweden\\
% ...
$^{d}$ Also at the Novosibirsk State University, Novosibirsk, 630090, Russia\\
% ...
     \vspace{0.4cm}
}
}

\begin{abstract}
We report the most precise measurements to date of the strong-phase parameters between $D^0$ and $\bar{D}^0$ decays to
$K^0_{S,L}\pi^+\pi^-$ using a sample of 2.93 fb$^{-1}$ of $e^+e^-$ annihilation data collected at a center-of-mass energy of 3.773 GeV with the BESIII detector at the BEPCII collider.  Our results provide the key inputs for a binned model-independent determination of the Cabibbo-Kobayashi-Maskawa angle $\gamma/\phi_3$ with $B$ decays. Using our results, the decay model sensitivity to the $\gamma/\phi_3$ measurement is expected to be between 0.7$^{\circ}$ and 1.2$^{\circ}$, approximately a factor of three smaller than that achievable with previous measurements, based on the studies of the simulated data.  The improved precision of this work ensures that measurements of $\gamma/\phi_3$ will not be limited by knowledge of strong phases for the next decade. Furthermore, our results provide critical input for other flavor-physics investigations, including charm mixing, other measurements of $CP$ violation, and the measurement of strong-phase parameters for other $D$-decay modes.

\end{abstract}

\pacs{13.25.Ft, 14.40.Lb, 14.65.Dw}

\maketitle

The mechanism of $CP$ violation in particle physics is of primary
importance because of its impact on cosmological baryogenesis and
matter-antimatter asymmetry in the universe.  In the standard model
(SM), $CP$ violation is studied by measuring the elements of the
Cabibbo-Kobayashi-Maskawa (CKM) matrix~\cite{CKM}, using the convenient
representation given by the unitarity triangle (UT) formed in the
complex plane.  The angle $\gamma$ (also denoted $\phi_3$) of the UT
is of particular interest since it is the only one that can be
extracted from tree-level processes, for which the contribution of
non-SM effects is expected to be very small. Therefore, measurement of
$\gamma$ provides a benchmark for the SM with minimal theoretical
uncertainty~\cite{jhep1401_051,epjc79_159}.  A precision measurement
of $\gamma$ is an essential ingredient in comprehensive testing of the
SM description of $CP$ violation and probing for evidence of new
physics.  Direct measurements of $\gamma$ have not yet achieved the
required precision, with a world-average value of
$\gamma=(73.5^{+4.2}_{-5.1})^{\circ}$~\cite{pdg18}, to be compared to
the indirect determination of
$\gamma=(65.8^{+1.0}_{-1.7})^{\circ}$~\cite{CKMfitter}.  These
different determinations deviate by 1.5$\sigma$.  It has been
predicted that new physics at the tree level could introduce a
deviation in $\gamma$ up to $4^{\circ}$~\cite{prd92_033002}, which is
close to the current experimental precision.  Achieving sub-degree
precision in the determination of $\gamma$ is clearly a top priority
for current and future flavor-physics experiments.

Generally, three methods had been suggested to measure $\gamma$ so far: GLW~\cite{PLB253_483,PLB265_172}, ADS~\cite{PRL78_3257,PRD63_036005}, and Dalitz(GGSZ)~\cite{prd68_054018} analyses.
One of the most sensitive decay channels for measuring $\gamma$ is
$B^-\rightarrow DK^-$ with $D\rightarrow
K^0_S\pi^+\pi^-$~\cite{prd68_054018}, where $D$ represents a
superposition of $D^0$ and $\bar{D}^0$ mesons.  (Throughout this
paper, charge conjugation is assumed unless otherwise explicitly
noted.)  The model-independent approach~\cite{epjc47_347} requires a
binned Dalitz plot analysis of the amplitude-weighted average cosine
and sine of the relative strong-phase ($\Delta \delta_D$) between
$D^0$ and $\bar{D}^0\rightarrow K^0_S\pi^+\pi^-$ to determine
$\gamma$.  These strong-phase parameters were first studied by the
CLEO collaboration using 0.82 fb$^{-1}$ of
data~\cite{prd80_032002,prd82_112006}.  The limited precision of
CLEO's results contributes a systematic uncertainty of approximately
4$^{\circ}$ to the $\gamma$ measurement~\cite{jhep08_176}, currently
the dominant systematic limitation in this determination.  In the
coming decades, the statistical uncertainties of measuring $\gamma$
will be greatly reduced by LHCb and Belle II, potentially to
$1^{\circ}$ or less.  The model-independent approach provides the most
precise stand-alone $\gamma$ measurement~\cite{jhep08_176}, and
therefore improved measurements of the $D$ strong-phase parameters are
essential in maximizing the precision of $\gamma$ from these future data
sets.

In this Letter, we use the model-independent approach of Ref.~\cite{epjc47_347} for the determination of the strong-phase parameters between $D^0$ and $\bar{D}^0\rightarrow K^0_{S,L}\pi^+\pi^-$.  More details are presented in a companion paper submitted to Physical Review D~\cite{PRD}.  Our data sample was collected from $e^+e^-$ annihilations at $\sqrt{s}=3.773$~GeV, just above the energy threshold for production of $D \bar{D}$ events.  At this energy we take advantage of unique quantum correlations afforded by production through the $\psi(3770)$ resonance.  The total integrated luminosity of our sample is 2.93 fb$^{-1}$~\cite{lumi}, 3.6 times that of the CLEO measurement.   The expected improvement in precision of the strong-phase parameters will significantly reduce the uncertainties of determinations of $\gamma$~\cite{plb718_43,jhep10_097,jhep06_131,jhep08_176,prd85_112014} that utilize $D \to K^0_{S,L} \pi^+\pi^-$.  Additionally, improved knowledge of these strong-phase parameters will have significant impact in other applications, including measurements of the CKM angle $\beta$ (also denoted $\phi_1$) through time-dependent analyses of $B^0\rightarrow Dh^0$~\cite{prd94_052004} (where $h$ is a light meson) and $B^0\rightarrow D\pi^+\pi^-$~\cite{jhep03_195}, as well as measurements of charm mixing and $CP$ violation~\cite{jhep10_185,jhep04_033,prd99_012007,prl122_231802}.

For this study we analyze the $D \rightarrow K^0_S \pi^+ \pi^-$ Dalitz plot phase space of $m^2_-$ vs. $m^2_+$, where $m^2_-$ and $m^2_+$ are the squared invariant masses  of the $K^0_S \pi^-$ and $K^0_S \pi^+$, respectively.  The phase space is partitioned into eight pairs of irregularly shaped bins following the three schemes defined in Ref.~\cite{prd82_112006}, which are divided according to regions of similar strong-phase difference $\Delta\delta_D$ or maximum sensitivity to $\gamma$ in the presence of negligible (significant) background; here these schemes are referred to as ``equal
$\Delta\delta_D$" and ``(modified) optimal", respectively.  The bin index $i$ ranges from $-8$ to $8$ (excluding 0), with the bins symmetric under the exchange $m^2_-\leftrightarrow m^2_+$ ($i\leftrightarrow -i$).  The strong-phase parameters are denoted $c_i$ and $s_i$, where $c_i$ is the amplitude-weighted average of cos$\Delta\delta_D$ in the $i$th region of the Dalitz plot ($\mathcal{D}_i$) and is given by
\begin{equation}
c_i=\frac{\int_{\mathcal{D}_i}|\mathcal{A}||\bar{\mathcal{A}}|{\rm cos}\Delta\delta_Dd\mathcal{D}}{\sqrt{\int_{\mathcal{D}_i}|\mathcal{A}|^2d\mathcal{D}\int_{\mathcal{D}_i}|\bar{\mathcal{A}}|^2d\mathcal{D}}},
\end{equation}
where $\mathcal{A}$ and $\bar{\mathcal{A}}$ are the amplitudes for $D^0\rightarrow
K^0_S\pi^+\pi^-$ and $\bar{D}^0\rightarrow K^0_S\pi^+\pi^-$, respectively.  The term $s_i$ is defined
analogously, with cos$\Delta\delta_D$ replaced by sin$\Delta\delta_D$.
Because the effects of charm mixing and $CP$ violation in the $D$
decay are negligible, we take $c_i=c_{-i}$ and $s_i=-s_{-i}$.  The
measurement involves studying the density of the correlated
$D\rightarrow K^0_S\pi^+\pi^-$ vs. $D\rightarrow K^0_{S,L}\pi^+\pi^-$
Dalitz plots, as well as decays of a $D$ meson tagged in a $CP$
eigenstate decaying to $K^0_{S,L}\pi^+\pi^-$.  The expected yields can
be expressed in terms of the parameters $K_i$, $c_i$ and $s_i$ for
$D^0\rightarrow K^0_S\pi^+\pi^-$, and $K^{\prime}_i$, $c^{\prime}_i$
and $s^{\prime}_i$ for $D\rightarrow K^0_L\pi^+\pi^-$, where
$K_i^{(\prime)}$ is determined from the distribution of the flavor-tagged
$D^0\rightarrow K^0_{S,L}\pi^+\pi^-$ decays across the bins of
the Dalitz plot as
$K_i^{(\prime)}=h_D\int_{\mathcal{D}_i}|\mathcal{A}|^2d\mathcal{D}$
and $h_D$ is a normalization factor.  Therefore, the strong-phase
parameters $c_i$, $s_i$, $c_i^{\prime}$, and $s_i^{\prime}$ can be
determined by minimizing the likelihood function constructed from the
observed and expected yields of these decays.

Details about the BESIII detector design and performance are provided
in Ref.~\cite{Ablikim:2009aa}.
To measure strong-phase parameters, we select ``single-tag'' (ST) and
``double-tag'' (DT) samples as listed in Table~\ref{tab:numST}.  STs
are $D$ mesons reconstructed from their daughter particles in one of
17 decay modes, of which four are flavor-specific, five are $CP$-even,
seven are $CP$-odd, and one ($K^0_S\pi^+\pi^-$) is $CP$-mixed.  Note
that we count $D\rightarrow \pi^+\pi^-\pi^0$ as a $CP$-even eigenstate
while explicitly correcting for its small $CP$-odd component
\cite{Malde:2015mha}.  DTs are events with an ST and a second $D$
meson reconstructed as either $K^0_S\pi^+\pi^-$ or $K^0_L\pi^+\pi^-$.
The $K^0_{L}$ mesons are not directly reconstructed and their presence
is inferred by partial reconstruction technique where one particle is
identified by the missing energy and mass in the event.  DTs are only
formed in combinations where there is a maximum of one unreconstructed
particle.



\begin{table}[tp!]
\caption{ Summary of ST yields ($N_{\rm ST}$) and DT yields for $K^0_{S,L}\pi^+\pi^-$ vs. various tags. The uncertainties are statistical only.
The tag modes of $\pi^+\pi^-\pi^0$, $K^0_S\eta^{\prime}$, $K^0_L\pi^0\pi^{0}$ and the partially-reconstructed $K^0_S\pi^+\pi^-$ events are used for the first time.
}
\begin{center}
\begin{tabular}
{l|ccc} \hline\hline Mode  & $N_{\rm ST}$  &  $N^{\rm DT}_{K^0_S\pi^+\pi^-}$   & $N^{\rm DT}_{K^0_L\pi^+\pi^-}$  \\
\hline
Flavor tags                                                    &                                  &                               &                               \\
$K^+\pi^-$                                                 &$549373\pm756$         &   $4740\pm71$        & ~~$9511\pm115$   \\
$K^+\pi^-\pi^0$                                         &$1076436\pm1406$     &   $5695\pm78$        & $11906\pm132$       \\
$K^+\pi^-\pi^-\pi^+$                                 &~~$712034\pm1705$  &   $8899\pm95$        & $19225\pm176$      \\
$K^+e^-\bar{\nu}_e$                                  &~~$458989\pm5724$  &   $4123\pm75$        &                               \\
$CP$-even tags                                             &                                  &                               &                               \\
$K^+K^-$                                                   &~~$57050\pm231$      &~~$443\pm22$       &$1289\pm41$            \\
$\pi^+\pi^-$                                               &~~$20498\pm263$      &~~$184\pm14$       &~~$531\pm28$         \\
$K^0_S\pi^0\pi^0$                                     &~~$22865\pm438$       &~~$198\pm16$      &~~$612\pm35$         \\
$\pi^+\pi^-\pi^0$                                       &$107293\pm716$          &~~$790\pm31$      & $2571\pm74$            \\
$K^0_L\pi^0$                                              &~$103787\pm7337$     &~~$913\pm41$       &                            \\
$CP$-odd tags                                               &                                   &                             &                             \\
$K^0_S\pi^0$                                             &~~$66116\pm324$      &~~$643\pm26$     &~~$861\pm46$       \\
$K^0_S\eta_{\gamma\gamma}$                    &~~~~$9260\pm119$   &~~~$89\pm10$     &~~$105\pm15$        \\
$K^0_S\eta_{\pi^+\pi^-\pi^0}$                    &~~$2878\pm81$         &~~$23\pm5$          &~~$40\pm 9$           \\
$K^0_S\omega$                                            &~~$24978\pm448$     &~~$245\pm17$       &~~$321\pm25$       \\
$K^0_S\eta^{\prime}_{\pi^+\pi^-\eta}$         &~~$3208\pm88$         &~~$24\pm6$          &~~$ 38\pm 8$         \\
$K^0_S\eta^{\prime}_{\gamma\pi^+\pi^-}$  &~~~$9301\pm139$     &~~~~$81\pm10$   &~~$120\pm14$        \\
$K^0_L\pi^0\pi^{0}$                                    &~~~$50531\pm6128$ &~~$620\pm32$     &                    \\
Mixed $CP$ tags                                             &                                   &                             &                             \\
$K^0_S\pi^+\pi^-$                                      &$188912\pm756$         &~~$899\pm31$      & $3438\pm72$           \\
$K^0_S\pi^+\pi^-_{\rm miss}$                    &                           &~~$224\pm17$  &                    \\
$K^0_S(\pi^0\pi^0_{\rm miss})\pi^+\pi^-$  &                          &~~$710\pm34$    &                    \\
\hline\hline
\end{tabular}
\label{tab:numST}
\end{center}
\end{table}

The selection and yield determination procedures of ST and DT candidates are described in the companion article~\cite{PRD} and are summarized below.
The ST yields, $N_{\rm ST}$, are listed in the second column of Table~\ref{tab:numST}.
The yields of DT candidates consisting of $K^0_S\pi^+\pi^-$ {\it vs.} fully
reconstructed final states are determined with a two-dimensional
unbinned maximum-likelihood fit to the ${\rm M}^{\rm sig}_{\rm BC}$
(signal) vs. ${\rm M}^{\rm tag}_{\rm BC}$ (tag) distribution. The DT
candidates with an undetectable neutrino or $K^0_L$ are reconstructed
by combining a $K^0_S\pi^+\pi^-$ candidate with the remaining charged
or neutral particles, that are assigned to the other $D$ decay. The
variable ${\rm U}_{\rm miss} = E_{\rm miss}-|\vec{p}_{\rm miss}|$ (for
$K^+e^-\bar{\nu}_e$) or missing-mass squared (${\rm M}^2_{\rm miss}$)
are calculated from the missing energy and momentum in the event.  To
reduce background contributions, events with excess neutral energy or
charged tracks are rejected.

\begin{figure}[tp!]
\begin{center}
   \begin{minipage}[t]{8.8cm}
   \includegraphics[width=\linewidth]{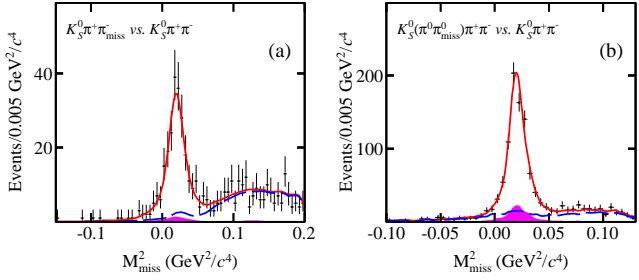}
   \end{minipage}
   \caption{ Fits to ${\rm M}_{\rm miss}^2$ distributions in
     data. Points with error bars are data, dotted (blue) curves
     are the fitted combinatorial backgrounds. The shaded areas (pink)
     show MC estimates of the peaking backgrounds mainly from (a)
     $D\rightarrow \pi^+\pi^-\pi^+\pi^-$ and (b) $D\rightarrow
     \pi^+\pi^-\pi^0\pi^0$, and the red solid curves are the total
     fits.}
\label{fig:MM2}
\end{center}
\end{figure}

The $K^0_{S}\pi^+\pi^-$ {\it vs.} $K^0_{S}\pi^+\pi^-$ DTs are crucial for
determining the $s_i$ values, and thus in order to increase the yield
for these events, we include two types of partially reconstructed
events, which more than doubles the yield.  The first
($K^0_S\pi^{\pm}\pi^{\mp}_{\rm miss}$) allows for one pion originating from
the $D$ meson to be unreconstructed in the detector.  For these
events, which have only three charged tracks recoiling against the
$D\rightarrow K^0_S\pi^+\pi^-$ ST, the missing pion is inferred from
the ${\rm M}^2_{\rm miss}$ of the event.  The second
($K^0_{S}(\pi^0\pi^0_{\rm miss})\pi^+\pi^-$) is the case where one
$K^0_S$ meson decays to $\pi^0\pi^0$, with only one $\pi^0$ detected while the other $\pi^0$ is undetected.
We select events with only two additional oppositely charged tracks
recoiling against the $D\rightarrow K^0_S\pi^+\pi^-$ ST and identify
these as the $\pi^+$ and $\pi^-$ from the other $D$ meson.  The
resulting distributions of ${\rm M}^2_{\rm miss}$ show clear signals
with minimal background, and signal yields are obtained with unbinned
maximum-likelihood fits, as is shown in Fig.~\ref{fig:MM2}.

\begin{figure}[tp!]
\begin{center}
   \begin{minipage}[t]{8.8cm}
   \includegraphics[width=\linewidth]{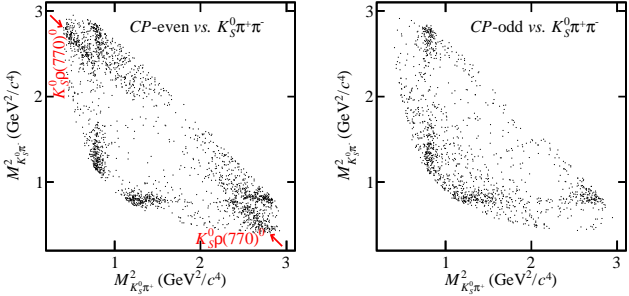}
   \end{minipage}
   \caption{Dalitz plots of $K^0_S\pi^+\pi^-$ events in data. The effect of the quantum correlation is clearly visible. The approximate locations of events from $K^0_S\rho(770)^0$ are indicated by arrows for clarity.}
\label{fig:Dalitz}
\end{center}
\end{figure}

The DT yields of $K^0_S\pi^+\pi^-$ and $K^0_L\pi^+\pi^-$ tagged by
different channels are shown in the third and fourth columns of
Table~\ref{tab:numST}, respectively.  Overall, the DT yields of $D \to
K^0_{S(L)}\pi^+\pi^-$ involving a $CP$ eigenstate are a factor of 5.3
(9.2) larger than those in Ref.~\cite{prd82_112006}, and the DT yields
of $K^0_S\pi^+\pi^-$ tagged with $D \to K^0_{S(L)}\pi^+\pi^-$ decays
are a factor of 3.9 (3.0) larger than those in
Ref.~\cite{prd82_112006}.  These increases come not only from the
larger data set available at BESIII but also from the additional tag
modes and the application of partial-reconstruction techniques.
Figure~\ref{fig:Dalitz} shows the Dalitz plots of $CP$-even and
$CP$-odd tagged $D\rightarrow K^0_S\pi^+\pi^-$ events selected in the
data.  The effect of quantum correlations arising from production
through $\psi(3770)\rightarrow D^0\bar{D}^0$ is demonstrated by the
differences between these plots.  Most noticeably, the $CP$-odd
component $K^0_S \rho(770)^0$ is visible in $CP$-even tagged
$K^0_S\pi^+\pi^-$ samples but absent from $CP$-odd samples.

The DT yield for the $i$th bin of the Dalitz plot of each tagged
$D\rightarrow K^0_{S(L)}\pi^+\pi^-$ sample, $N^{\rm obs}_{i}$, can be
determined by fitting the DT events observed in this bin. Here the
yield includes the signal and any peaking background component. The
expected DT yields in the $i$th bin of Dalitz plot of each tagged
$D\rightarrow K^0_{S(L)}\pi^+\pi^-$ sample, $N^{\rm exp}_{i}$, are
sums of the expected signal yields and the expected peaking
backgrounds.  It should be noted that detector resolution effects can
cause individual events to migrate between Dalitz plot bins after
reconstruction. Such migration effects vary among bins due to the
irregular bin shapes, coupled with the rapid variations of the Dalitz
plot density.  Furthermore, migrations differ between $D\rightarrow
K^0_S\pi^+\pi^-$ and $D\rightarrow K^0_L\pi^+\pi^-$ decays due to
different resolutions in the Dalitz plots (0.0068 GeV$^2/c^4$ for
$D\rightarrow K^0_S\pi^+\pi^-$ and 0.0105 GeV$^2/c^4$ for
$D\rightarrow K^0_L\pi^+\pi^-$).  The resultant bin migrations range
within (3-12)\% and (3-18)\% for the $K^0_S\pi^+\pi^-$ and
$K^0_L\pi^+\pi^-$ signals, respectively.  Therefore, in the determination
of the DT yields, simulated efficiency matrices are introduced to
account for bin migration and reconstruction efficiencies~\cite{PRD}.
Studies indicate that neglecting bin migration introduces biases in
the determination of $c_i~(s_i)$ that average a factor of 0.7 (0.3)
times the statistical uncertainty of this analysis, so it is important
to correct for this effect.  The values of $K_i$ and $K_i^{\prime}$
that are used to evaluate $N^{\rm exp}_{i}$ are determined from the
flavor-tagged DT yields, where corrections from doubly
Cabibbo-suppressed decays, efficiency and migration effects have been
applied, which are explained in detail in Ref.~\cite{PRD}.

The values of $c^{(\prime)}_i$ and $s^{(\prime)}_i$ are obtained by minimizing the negative log-likelihood function constructed as
\begin{eqnarray}
-2{\rm log}\mathcal{L}&=&-2\sum\limits_{i}\sum\limits_{j} {\rm ln}P(N^{\rm obs}_{ij},\langle N^{\rm exp}_{ij}\rangle)_{K^0_S\pi^+\pi^-,K^0_{S(L)}\pi^+\pi^-} \nonumber \\
                                  &&-2\sum\limits_{i} {\rm ln}P(N^{\rm obs}_i,\langle N_i^{\rm exp}\rangle)_{CP,K^0_{S(L)}\pi^+\pi^-} +\chi^2, \nonumber
\label{eq:likelihood}
\end{eqnarray}
where $P(N^{\rm obs},\langle N^{\rm exp} \rangle)$ is the Poisson
probability to observe $N^{\rm obs}$ events given the expected number
$\langle N^{\rm exp} \rangle$.  Here the sums are over the bins of the
$D^0\to K^{0}_{S(L)}\pi^+\pi^-$ Dalitz plots.  The $\chi^2$ term is
used to constrain the difference $c_i^{\prime}-c_i$
($s_i^{\prime}-s_i$) to the predicted quantity $\Delta c_i$ ($\Delta
s_i$). The values of $\Delta c_i$ and $\Delta s_i$ are estimated based
on the decay amplitudes of $D^0\rightarrow
K^0_S\pi^+\pi^-$~\cite{prd98_112012} and $D^0\rightarrow
K^0_L\pi^+\pi^-$, where the latter is constructed by adjusting the
$D^0\rightarrow K^0_S\pi^+\pi^-$ model taking the
$K^0_S$ and $K^0_L$ mesons to have opposite $CP$, as is discussed in
Refs.~\cite{prd80_032002,prd82_112006}.  The details of assigning
$\Delta c_i$ ($\Delta s_i$) and their uncertainties $\delta\Delta c_i$
($\delta\Delta s_i$) are presented in Table VI of Ref.~\cite{PRD}.

\begin{figure*}[tp!]
\begin{center}
   \includegraphics[width=1.0\linewidth]{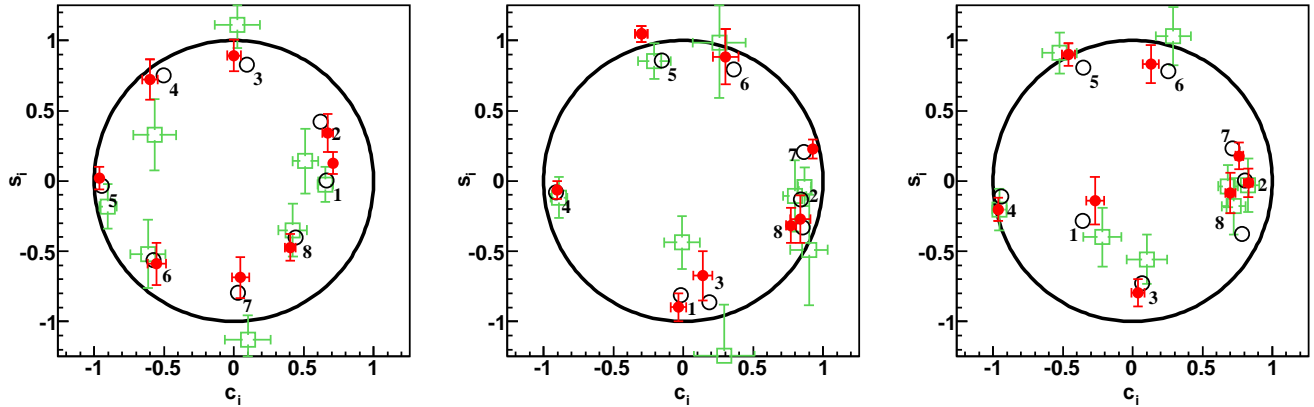}
   \caption{ The $c_i$ and $s_i$ measured in this work (red dots with error bars), the predictions of Ref.~\cite{prd98_112012} (black open circles) and the results of Ref.~\cite{prd82_112006} (green open squares with error bars).  The left, middle and right plots are from the equal $\Delta\delta_D$, optimal and modified optimal binnings, respectively.  The circle indicates the boundary of the physical region $c^{2}_i+s^{2}_i=1$.
   }
\label{fig:cisi}
\end{center}
\end{figure*}

The measured strong-phase parameters $c^{(\prime)}_i$ and
$s^{(\prime)}_i$ are presented in Fig.~\ref{fig:cisi} and
Table~\ref{tab:cisifit_final}.
The estimation of systematic uncertainties is described in detail in Ref.~\cite{PRD}.  In addition
to our results, Fig.~\ref{fig:cisi} includes the predictions of
Ref.~\cite{prd98_112012} and the results from
Ref.~\cite{prd82_112006}, which show reasonable agreement.

In summary, measurements of the strong-phase parameters between $D^0$ and $\bar{D}^0\rightarrow K^0_{S,L}\pi^+\pi^-$ in bins of phase space have been performed using 2.93 fb$^{-1}$ of data collected at $\sqrt{s}$=3.773 GeV with the BESIII detector.
Compared to the previous CLEO measurement~\cite{prd82_112006}, two main improvements have been incorporated.   First, additional tag decay modes are used. In particular the inclusion of the
$\pi^+\pi^-\pi^0$ tag improves the sensitivity to $c_i$ and the addition of the $K^0_{S}(\pi^0\pi^0_{\rm miss})\pi^+\pi^-$ improves the sensitivity to $s_i$. Second, corrections for bin migration have been included, as their neglect would lead to uncertainties comparable to the statistical uncertainty. The results presented in this Letter are on average a factor of 2.5 (1.9) more precise for $c_i$ ($s_i$) and a factor of 2.8 (2.2) more precise for $c^{\prime}_i$ ($s^{\prime}_i$) than has been achieved previously.  The strong-phase parameters provide an important input for a wide range of $CP$ violation measurements in the beauty and charm sectors, and also for measurements of strong-phase parameters in other $D$ decays where $D \to K^0_S\pi^+\pi^-$ is used as a tag~\cite{plb747_9,plb757_520,jhep01_082,jhep01_144,prd85_092016}.

To assess the impact of our $c_i$ and $s_i$ results on a measurement of $\gamma$, we use a large simulated data set of $B^-\rightarrow DK^-$, $D\rightarrow K^0_S\pi^+\pi^-$ events.
Based on the MC simulation, the uncertainty in $\gamma$ associated with our uncertainties for $c_i$ and $s_i$ is found to be 0.7$^{\circ}$, 1.2$^{\circ}$ and 0.8$^{\circ}$ for the equal $\Delta\delta_D$, optimal and modified optimal binning schemes, respectively.
For comparison, the corresponding results from CLEO are 2.0$^{\circ}$, 3.9$^{\circ}$ and 2.1$^{\circ}$~\cite{prd82_112006}.
Therefore, the uncertainty on $\gamma$ arising from
knowledge of the charm strong phases is approximately a factor of three smaller than was possible with the CLEO measurements.
For the first time, the dominant systematic uncertainty for $\gamma$ measurement from the strong-phase parameters will be constrained to around  $1^{\circ}$, or less, for $\gamma$ measurements with future $B$ experiments~\cite{plb718_43,jhep10_097,jhep06_131,jhep08_176,prd85_112014}.  The predicted statistical uncertainties on $\gamma$ from LHCb prior to the start of High-Luminosity LHC operation in the mid 2020s, and from Belle II are expected to be around 1.5$^\circ$~\cite{1808.08865,1808.10567}.  The improved precision achieved here will ensure that measurements of $\gamma$ from LHCb and Belle II over the next decade are not limited by the knowledge of these strong-phase parameters.

These strong-phase parameters also provide critical inputs in model-independent measurements of charm-mixing and $CP$ violation in $D^0\rightarrow K^0_S\pi^+\pi^-$
decays~\cite{prd99_012007,prl122_231802}.
As detailed in Ref.~\cite{prd99_012007}, the precision of the charm-mixing parameters $x$ and $y$ is dependent on $c_i$ and $s_i$ inputs. With $5\times10^8$ $D^0\rightarrow K^0_S\pi^+\pi^-$ signal decays, which is the anticipated yield at LHCb in 2030, the uncertainty from the CLEO determination of the strong phases is expected to be approximately a factor 3.8 (5.0) larger than the statistical uncertainty for $x$ ($y$)~\cite{prd99_012007}, leading to measurements where the overall precision is limited by the strong-phase inputs.
To evaluate the impact of our $c_i$ and $s_i$ results on the measurements of $x$ and $y$, we generate $5\times10^8$ $D^0\rightarrow K^0_S\pi^+\pi^-$ signal decays using input
charm-mixing parameters $x=0.4\%$ and $y=0.6\%$, with no $CP$ violation. By using the ``bin-flip method"~\cite{prd99_012007} and keeping the $c_i$ and $s_i$ constrained according to
our measurements, the expected statistical uncertainties on $x$ and $y$ are $0.027\%$ and $0.061\%$, respectively. Thus, compared with the expected statistical uncertainties on $x$
(0.034\%) and $y$ (0.091\%) with CLEO inputs~\cite{prd99_012007}, it is clear that our results will significantly reduce uncertainties on future charm-mixing measurements.
\begin{table*}
\caption{The measured strong-phase parameters $c^{(\prime)}_i$ and $s^{(\prime)}_i$, where the first uncertainties are statistical, including that related to the $\Delta c_i$ and $\Delta s_i$ constraints,} and the second are systematic.
\begin{center}
\vspace{-0.2cm}
\resizebox{!}{3.45cm}{
\begin{tabular}{crr|rr|rr} \hline
\hline
    & \multicolumn{2}{c|}{\normalsize Equal $\Delta\delta_D$ binning} & \multicolumn{2}{c|}{\normalsize Optimal binning} & \multicolumn{2}{c}{\normalsize Modified optimal binning} \\
    & {\normalsize $c_i$}~~~~~~~~~~~~~~~~~  & {\normalsize $s_i$}~~~~~~~~~~~~~~~~~
    & {\normalsize $c_i$}~~~~~~~~~~~~~~~~~  & {\normalsize $s_i$}~~~~~~~~~~~~~~~~~
    & {\normalsize $c_i$}~~~~~~~~~~~~~~~~~  & {\normalsize $s_i$}~~~~~~~~~~~~~~~~~ \\
\hline
\normalsize
1 & $~~0.708(0.020)(0.009)$ & $~~0.128(0.076)(0.017)$ & $ -0.034(0.052)(0.017)$ & $ -0.899(0.094)(0.030)$ & $ -0.270(0.061)(0.019)$ & $ -0.140(0.168)(0.027)$ \\
2 & $~~0.671(0.035)(0.016)$ & $~~0.341(0.134)(0.015)$ & $~~0.839(0.062)(0.037)$ & $ -0.272(0.166)(0.031)$ & $~~0.829(0.027)(0.018)$ & $ -0.014(0.100)(0.018)$ \\
3 & $~~0.001(0.047)(0.019)$ & $~~0.893(0.112)(0.019)$ & $~~0.140(0.064)(0.028)$ & $ -0.674(0.172)(0.037)$ & $~~0.038(0.044)(0.021)$ & $ -0.796(0.095)(0.020)$ \\
4 & $ -0.602(0.053)(0.016)$ & $~~0.723(0.143)(0.015)$ & $ -0.904(0.021)(0.009)$ & $ -0.065(0.062)(0.006)$ & $ -0.963(0.020)(0.009)$ & $ -0.202(0.080)(0.014)$ \\
5 & $ -0.965(0.019)(0.013)$ & $~~0.020(0.081)(0.009)$ & $ -0.300(0.042)(0.013)$ & $~~1.047(0.055)(0.014)$ & $ -0.460(0.044)(0.011)$ & $~~0.899(0.078)(0.013)$ \\
6 & $ -0.554(0.062)(0.024)$ & $ -0.589(0.147)(0.030)$ & $~~0.303(0.088)(0.027)$ & $~~0.884(0.191)(0.042)$ & $~~0.130(0.055)(0.017)$ & $~~0.832(0.131)(0.029)$ \\
7 & $~~0.046(0.057)(0.023)$ & $ -0.686(0.143)(0.028)$ & $~~0.927(0.016)(0.008)$ & $~~0.228(0.066)(0.015)$ & $~~0.762(0.025)(0.012)$ & $~~0.178(0.094)(0.016)$ \\
8 & $~~0.403(0.036)(0.017)$ & $ -0.474(0.091)(0.027)$ & $~~0.771(0.032)(0.015)$ & $ -0.316(0.123)(0.020)$ & $~~0.699(0.035)(0.012)$ & $ -0.085(0.141)(0.018)$ \\
\hline
    & {\normalsize $c^{\prime}_i$}~~~~~~~~~~~~~~~~~  & {\normalsize $s^{\prime}_i$}~~~~~~~~~~~~~~~~~
    & {\normalsize $c^{\prime}_i$}~~~~~~~~~~~~~~~~~  & {\normalsize $s^{\prime}_i$}~~~~~~~~~~~~~~~~~
    & {\normalsize $c^{\prime}_i$}~~~~~~~~~~~~~~~~~  & {\normalsize $s^{\prime}_i$}~~~~~~~~~~~~~~~~~ \\
\hline
1 & $~~0.801(0.020)(0.013)$ & $~~0.137(0.078)(0.016)$ & $~~0.240(0.054)(0.015)$ & $ -0.854(0.106)(0.032)$ & $ -0.198(0.067)(0.025)$ & $ -0.209(0.181)(0.027)$\\
2 & $~~0.848(0.036)(0.016)$ & $~~0.279(0.137)(0.016)$ & $~~0.927(0.054)(0.036)$ & $ -0.298(0.162)(0.029)$ & $~~0.945(0.026)(0.018)$ & $ -0.019(0.100)(0.017)$\\
3 & $~~0.174(0.047)(0.016)$ & $~~0.840(0.118)(0.020)$ & $~~0.742(0.060)(0.030)$ & $ -0.350(0.180)(0.039)$ & $~~0.477(0.040)(0.019)$ & $ -0.709(0.119)(0.028)$\\
4 & $ -0.504(0.055)(0.019)$ & $~~0.784(0.147)(0.014)$ & $ -0.930(0.023)(0.019)$ & $ -0.075(0.075)(0.007)$ & $ -0.948(0.021)(0.013)$ & $ -0.235(0.086)(0.014)$\\
5 & $ -0.972(0.021)(0.017)$ & $ -0.008(0.089)(0.009)$ & $ -0.173(0.043)(0.010)$ & $~~1.053(0.062)(0.016)$ & $ -0.359(0.046)(0.011)$ & $~~0.943(0.084)(0.013)$\\
6 & $ -0.387(0.069)(0.025)$ & $ -0.642(0.152)(0.033)$ & $~~0.554(0.073)(0.032)$ & $~~0.605(0.184)(0.042)$ & $~~0.333(0.051)(0.019)$ & $~~0.701(0.137)(0.028)$\\
7 & $~~0.462(0.056)(0.019)$ & $ -0.550(0.159)(0.030)$ & $~~0.975(0.017)(0.008)$ & $~~0.198(0.071)(0.014)$ & $~~0.878(0.026)(0.015)$ & $~~0.188(0.098)(0.016)$\\
8 & $~~0.640(0.036)(0.015)$ & $ -0.399(0.099)(0.026)$ & $~~0.798(0.035)(0.017)$ & $ -0.253(0.141)(0.019)$ & $~~0.740(0.037)(0.014)$ & $ -0.025(0.149)(0.019)$\\
\hline\hline
\end{tabular}
}
\label{tab:cisifit_final}
\end{center}
\end{table*}

The BESIII collaboration thanks the staff of BEPCII and the IHEP computing center for their strong support. This work is supported in part by National Key Basic Research Program of China under Contract No. 2015CB856700; National Natural Science Foundation of China (NSFC) under Contracts Nos. 11625523, 11635010, 11735014, 11775027, 11822506, 11835012; the Chinese Academy of Sciences (CAS) Large-Scale Scientific Facility Program; Joint Large-Scale Scientific Facility Funds of the NSFC and CAS under Contracts Nos. U1532257, U1532258, U1732263, U1832207; CAS Key Research Program of Frontier Sciences under Contracts Nos. QYZDJ-SSW-SLH003, QYZDJ-SSW-SLH040; 100 Talents Program of CAS; INPAC and Shanghai Key Laboratory for Particle Physics and Cosmology; ERC under Contract No. 758462; German Research Foundation DFG under Contracts Nos. Collaborative Research Center CRC 1044, FOR 2359; Istituto Nazionale di Fisica Nucleare, Italy; Koninklijke Nederlandse Akademie van Wetenschappen (KNAW) under Contract No. 530-4CDP03; Ministry of Development of Turkey under Contract No. DPT2006K-120470; National Science and Technology fund; STFC (United Kingdom); The Knut and Alice Wallenberg Foundation (Sweden) under Contract No. 2016.0157; The Royal Society, UK under Contracts Nos. DH140054, DH160214; The Swedish Research Council; U. S. Department of Energy under Contracts Nos. DE-FG02-05ER41374, DE-SC-0010118, DE-SC-0012069; University of Groningen (RuG) and the Helmholtzzentrum fuer Schwerionenforschung GmbH (GSI), Darmstadt; This paper is also supported by Beijing municipal government under Contract No CIT\&TCD201704047, and by the Royal Society under Contract No. NF170002.


\end{document}